\documentclass[twocolumn,preprintnumbers,amsmath,amssymb]{revtex4}
\input{epsf}

\newcommand{\be}{\begin{equation}}
\newcommand{\ee}{\end{equation}}
\newcommand{\bea}{\begin{eqnarray}}
\newcommand{\eea}{\end{eqnarray}}
\newcommand{\bef}{\begin{figure}}
\newcommand{\eef}{\end{figure}}

\newcommand{\simge}{\,{}^>_{\sim}\,}
\newcommand{\simle}{\,{}^<_{\sim}\,}

\def\h#1{$^{#1}$H}
\def\he#1{$^{#1}$He}
\def\li#1{$^{#1}$Li}
\def\be#1{$^{#1}$Be}

\def\eps@scaling{0.96}

\def\showone#1{
  \centering
  \leavevmode
  \epsfxsize=\eps@scaling\linewidth
  \epsfbox{#1.eps}
}

\def\epstwo@scaling{0.48}

\def\showtwo#1#2{
  \centering
  \leavevmode
  \epsfxsize=\epstwo@scaling\linewidth
  \epsfbox{#1.eps} \hfil
  \epsfxsize=\epstwo@scaling\linewidth
  \epsfbox{#2.eps}
}

\begin{document}

\title{
Bounds on long-lived
charged massive particles from Big Bang nucleosynthesis}
\author{Karsten Jedamzik} 
\affiliation{Laboratoire de Physique Math\'emathique et Th\'eorique, C.N.R.S.,
Universit\'e de Montpellier II, 34095 Montpellier Cedex 5, France}

\begin{abstract}
The Big Bang nucleosynthesis (BBN)
in the presence of charged massive particles
(CHAMPs) is studied in detail. All currently known effects due to the
existence of bound states between CHAMPs and nuclei, including possible
late-time destruction of \li6 and \li7 are included.
The study sets conservative bounds on CHAMP abundances in the
decay time range $3\times 10^2{\rm sec}\simle\tau_x\simle 10^{12}{\rm sec}$.
It is stressed that
the production of \li6 at early times $T\sim 10\,$keV
is overestimated by a factor
$\sim 10$ when the approximation of the Saha equation for the \he4
bound state fraction is utilised.
To obtain conservative limits on the abundance of CHAMPs, 
a Monte-Carlo analysis with  
$\sim 3\times 10^6$ independent BBN runs, varying
reaction rates of nineteen different reactions, 
is performed (see attached erratum, however).
The analysis yields the surprising result that except for small areas in the
particle parameter space conservative 
constraints on the abundance of
decaying charged particles are currently
very close to those of neutral
particles. It is shown that, in case the rates of a number of heretofore
unconsidered reactions may be determined
reliably in future, it is conceivable that
the limit on CHAMPs in the early Universe could be tightened by orders
of magnitude. An {\bf ERRATUM} gives limits on primordial CHAMP densities 
when the by Ref.~\cite{Kamimura:2008fx} recently more accurately 
determined CHAMP reaction rates are employed.  
\end{abstract}


\maketitle

\section{Introduction}

Big Bang nucleosynthesis (BBN) has proven itself as a powerful tool in
constraining the conditions of the early Universe and physics beyond the
standard model. Thus bounds on a variety of hypothesis have been derived
including, for example, modifications of gravity,
baryon inhomogeneity, matter-antimatter domains, 
non-zero lepton chemical potentials, and relic decaying particles.
It has been recently realized that BBN may also place bounds
on the abundance of charged, weak-scale mass particles existing
during 
and after BBN. Though it had been noted already earlier that 
negatively charged,
weak-scale mass particles (CHAMPs)
form bound-states with positively charged nuclei
towards the end of 
BBN~\cite{boundstate}, it has only recently been put forward that
this may impact considerably
the light-element yields synthesized during 
BBN~\cite{Pospelov,Kohri,Kaplinghat}.  

Here the most important proposed change is due to a 
catalysm of reactions
such as \h2 + \he4 $\to$ \li6 $+\gamma$~\cite{Pospelov}. 
Other less important
modifications, concerning constraints on CHAMPs, had also been 
considered~\cite{Kaplinghat,Bird,Jittoh}. 
Being of quadrupole (E2)
nature the \h2 + \he4 reaction has a very small rate (S-factor: $10^{-8}$MeV
barn) thus yielding typically very little \li6/\h1$\sim 10^{-14}$ in standard
BBN. When the helium-nuclei is in a bound state (\he4-$X^-$) the above
reaction may be replaced by its photonless analogue: 
\h2 + (\he4-$X^-$) $\to$ \li6 $+X^-$, with a cross section  
estimated orders of magnitude larger than that for the standard BBN
\li6-synthesizing process. 
Initial estimates for this enhancemant factor were given at around
$6\times 10^7$~\cite{Pospelov,Cyburt}. It was argued that since
$\sim 1$ of all $X^-$ are within bound states with \he4 at temperatures
$T\simle 8\,$keV (cf. Fig.~\ref{fig29}) very small abundances of
CHAMPs present at $t\approx 10^4$sec in the early Universe
could already overproduce the \li6 isotope with respect to 
observations~\cite{li6}. Assuming a (too restrictive) \li6/\h1 
$\simle 2\times 10^{-11}$ constraint,
bounds as strong as $n_{X^-}/s\simle 2.5\times 10^{-17}$,
the CHAMP-to-entropy ratio, were derived. These bounds were subsequently
weakened by one order of magnitude when a more proper evaluation of the
rate for the \h2 + (\he4-$X^-$) $\to$ \li6 $+X^-$ process~\cite{Hamaguchi} 
was derived. 
Such bounds have now been utilised by a number of groups
to constrain, for example, abundances of supersymmetric staus 
$\tilde{\tau}$s in the early 
Universe~\cite{Pospelov,Cyburt,Hamaguchi,Kawasaki,Pradler}.

Recently, I have shown that there are several changes to the "naive" picture
of synthesis of \li6 in the presence of CHAMPs~\cite{jeda6}. First, 
(\he4-$X^-$) bound
states may be destroyed during the electromagnetic cascades induced by the
decay of CHAMPs, rendering \li6 production in some parts of parameter
space much less efficient.
More importantly,
when CHAMPs are still present at times $t\simge 10^6$sec 
Big Bang nucleosynthesis enters a second phase of Coulomb-unsupressed
reactions on bound states between charge $Z=1$ nuclei and a CHAMP. 
It has been shown that reactions such as 
\li6 + (\h1-$X^-$)$\to$ \he4 + \he3 +$X^-$, 
\be7 + (\h1-$X^-$)$\to$ $^8$B +$X^-$, etc, are capable of completely
destroying any priorly synthesized \li6 and \li7. This is possible
in particular at somewhat higher CHAMP-to-entropy $Y_X=n_{X}/s$ ratios.
In order to estimate the efficiency of such destruction, not only was
it required to estimate the cross sections for such \li6 and \li7 destroying
reactions, but also those of CHAMP exchange reactions such as
(\h1-$X^-$)+ \h2$\to$\h1 + (\h2-$X^-$), capable of significantly reducing
the \h1 bound state fraction. Altogether nineteen reactions of significant
importance for the late-time nucleosynthesis $t\simge 10^6$sec have been
identified. The rates for all these reactions were determined in the Born
approximation. Concerning details on the
BBN with CHAMPs at late times, the importance of particular
reactions, and their evaluation, the reader is referred to the original
paper~\cite{jeda6}. Unfortunately, the Born approximation is not a particular
good approximation for determining rates of these nineteen important
reactions for CHAMP BBN, leaving significant uncertainty in the BBN yields
with late decaying $\tau_x\simge 10^6$sec CHAMPs. 

Given the above, it seems very premature to rule out CHAMPs in the early
Universe simply by their production of \li6 at $T\approx 8\,$keV. In this
letter constraints on the abundances on CHAMPs are derived
which take full account of all the above
mentioned extra physics priorly neglected. 
Here constraints
will be placed for two different decay time regimes, for 
$3\times 10^2{\rm sec }\,\simle \tau_x\simle\, 5\times 10^5{\rm sec}$ in Section 2, where
the important rates are relativily well known, and for
$5\times 10^5{\rm sec }\,\simle \tau_x\simle\, 10^{12}{\rm sec}$ in Section 3, where
a Monte-Carlo analysis is employed to derive conservative limits. It will
be seen that constraints change by large
factors with respect to those priorly given, particularly for long CHAMP
decay times.

\section{Constraints on CHAMPs with intermediate life times}

\bef
\epsfxsize=8.5cm
\epsffile[50 50 410 302]{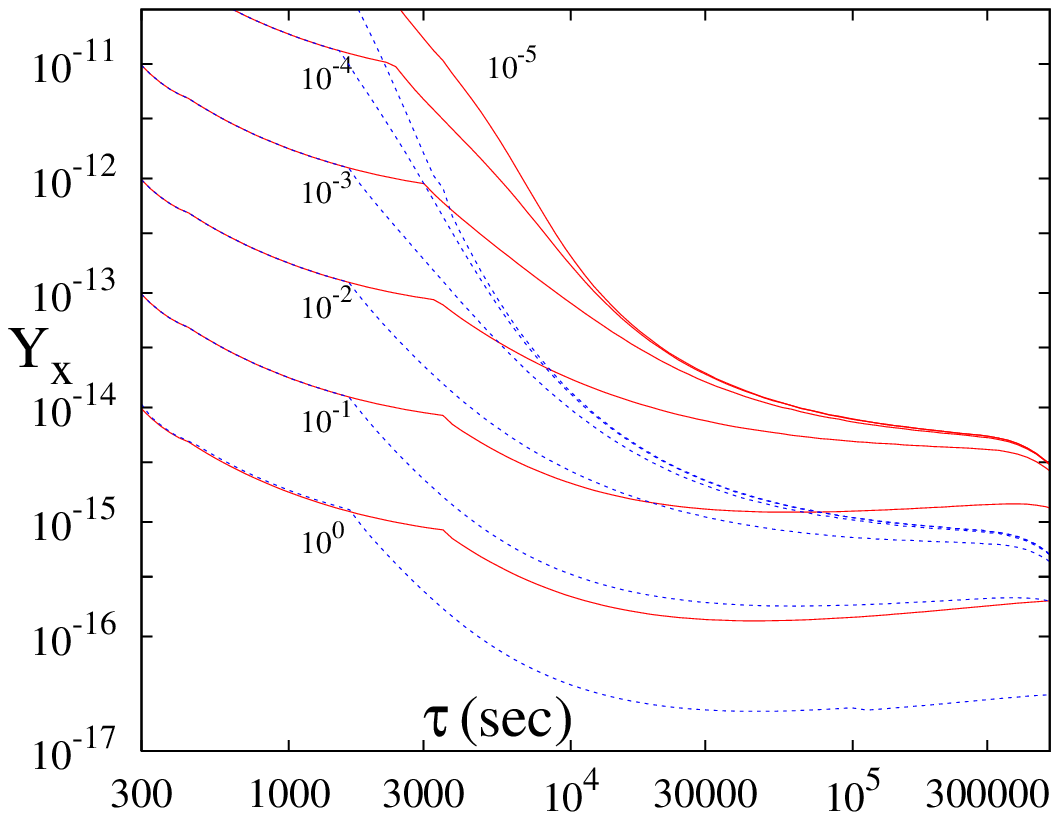}
\caption{Limits on the primordial CHAMP-to-entropy ratio 
$Y_x = n_{X}/s$ (with $n_{X^-}/s = Y_x/2$)
for CHAMPs with intermediate life times. Shown are constraint
lines for CHAMPs of mass $M_x =1\,$TeV and a variety
of hadronic branching ratios $B_h = 10^{-5}-1$,
as labeled in the figure. Solid (red)
lines correspond to the conservativelimit \li6/\li7 $< 0.66$, whereas dashed (blue) lines correspond to
\li6/\li7 $< 0.1$. It is seen that only for CHAMPs with $B_h\simle 10^{-2}$
the effects of bound states become important. For smaller decay times
$\tau_x$ the limits on CHAMP abundances are virtually identical to those
on the abundance of neutral relic decaying particles~\cite{jeda5}.}
\label{fig31}
\eef

\bef
\epsfxsize=8.5cm
\epsffile[50 50 410 302]{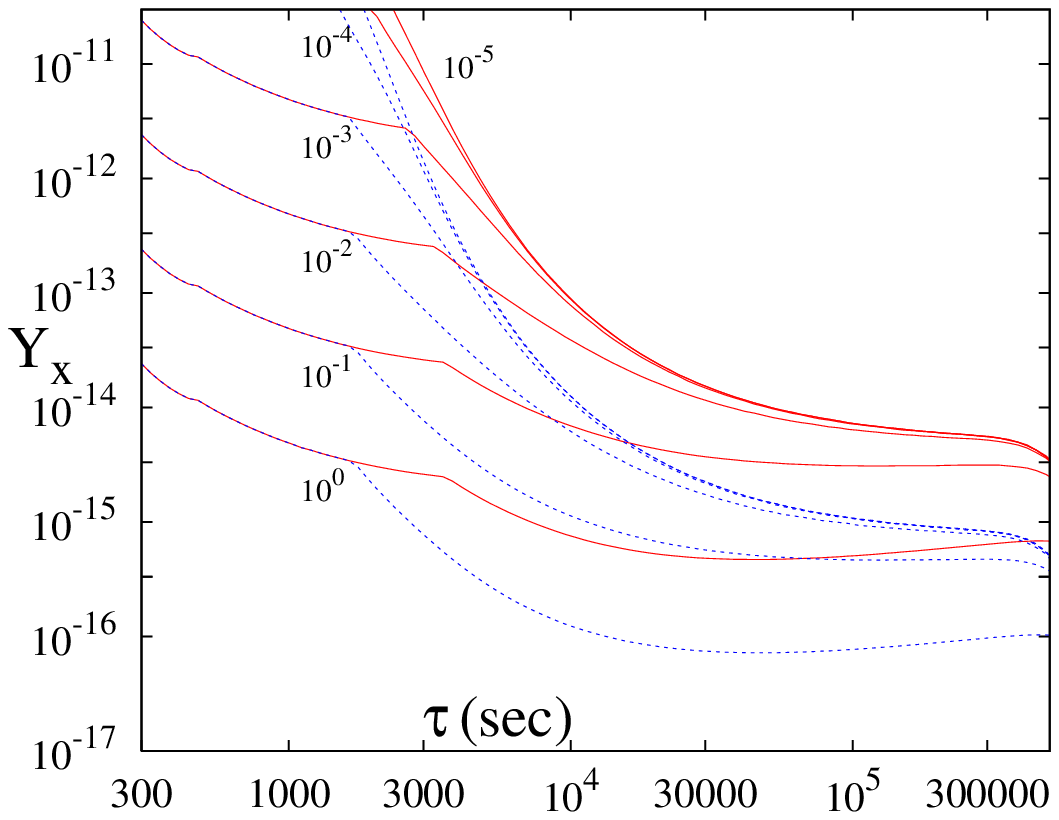}
\caption{Same as Fig.~\ref{fig31}, but for $M_x =100\,$GeV.}
\label{fig41}
\eef

\bef
\epsfxsize=8.5cm
\epsffile[50 50 410 302]{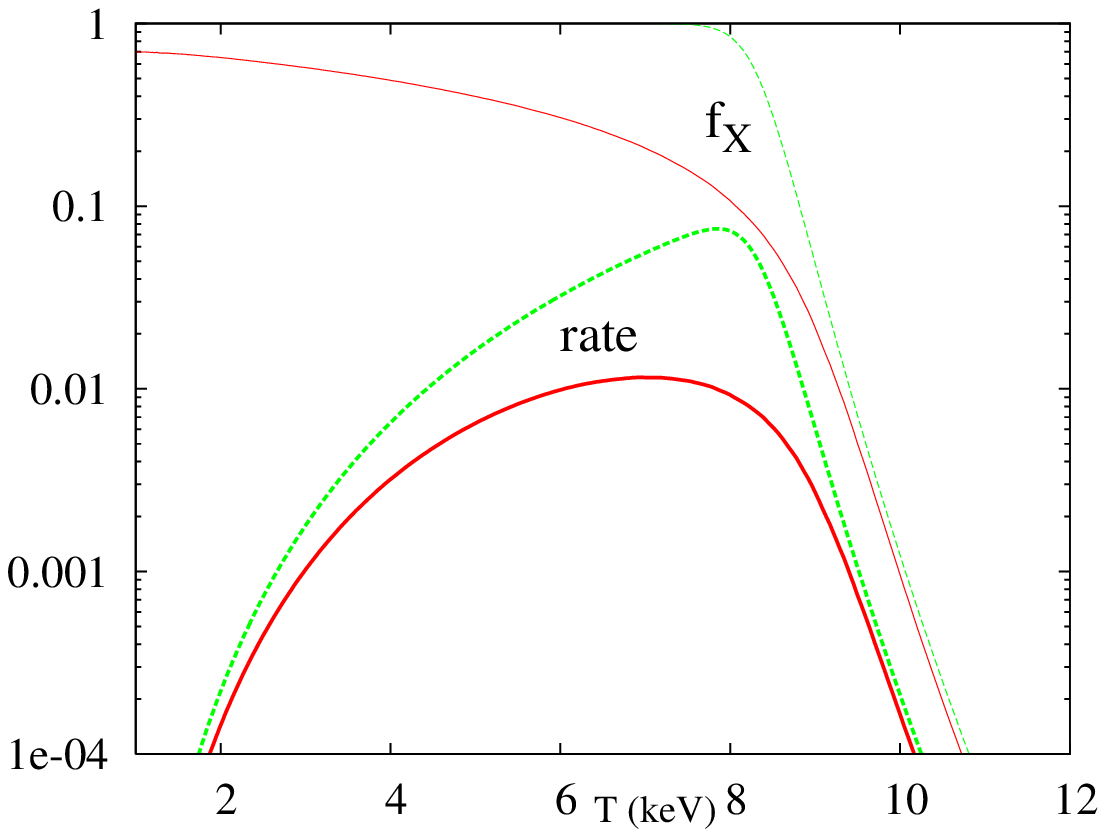}
\caption{Fraction of CHAMPs $f_X$ which are bound to \he4 as a function
of temperature for (a) the approximation by the Saha equation, thin-dashed (green) and (b) full numerical integration of the rate equation, thin-solid (red).
Also shown is the product of $f_X$ with the rate 
$\langle\sigma v\rangle$ for the \li6 producing reaction 
\h2(\he4$-X^- ,X^-){}^6$Li, simply
denoted as ``rate'' and in arbritrary units, for
both cases (a) thick-dashed (green) and (b) thick-solid (red). The figure
illustrates that \li6 production at $T\approx 8\,$keV due to bound states
is overestimated by a factor $\sim 10$ when the approximation of the Saha
equation is utilised.}
\label{fig29}
\eef

In the next two sections constraints from BBN on the existence of CHAMPs in the
early Universe are presented. Such constraints get increasingly more
uncertain as the life time $\tau_x$ of a CHAMP increases. For 
$\tau_x\simle 3\times 10^2$sec CHAMPs have almost no impact on BBN beyond
those of their injection of electromagnetically and hadronically interacting
particles during their decay~\cite{remark301}. Such constraints have already
been discussed in detail in the literature and the reader is referred 
to, for example, Ref.~\cite{jeda5} for details. 
For $3\times 10^2{\rm sec}\simle \tau_x\simle 
5\times 10^5{\rm sec}$ constraints are still fairly
reliable and depend mostly on the \h2(\he4$-X^- ,X^-){}^6$Li rate.
Since this one has been determined beyond the Born 
approximation~\cite{Hamaguchi} and is most
likely known to within a factor three, limits should also be known 
up to such a factor. It has been found numerically that other reactions, 
such as\h3(\he4$-X^- ,X^-){}^7$Li
and \he3(\he4$-X^- ,X^-){}^7$Be play less of a role in setting constraints
at early times, even in case their rates significantly exceed those determined
in the Born approximation. 

When deriving constraints the following conservative
observationally determined limits on the light element abundances are
adopted:
\begin{eqnarray}
Y_p < 0.258 \nonumber \\
1.2\times 10^{-5} < {\rm ^2H/H} < 5.3\times 10^{-5} \nonumber \\
{\rm ^3He/^2H} < 1.52 \\
\label{obs}
8.5\times 10^{-11} < {\rm ^7Li/H} < 5\times 10^{-10} \nonumber \\
{\rm ^6Li/^7Li} < 0.66\, (0.1)\nonumber
\end{eqnarray}
Here $Y_p$ denotes the helium mass fraction. The observations
behind these limits are discussed
in further detail in Ref.~\cite{jeda5}. It should be noted here that
the frequently used bound \li6/\h1 $\simle 2\times 10^{-11}$ is 
too stringent. \li6 may be destroyed during
the life time of a Population II star. In fact, if the current 
discrepancy between
standard BBN predicted \li7/\h1 $\approx 4-5\times 10^{-10}$ and
Pop II star observed \li7/\h1 $\approx 1-2.5\times 10^{-10}$ is resolved
by factor $2-3$ stellar \li7 destruction, 
as claimed for example in Ref.~\cite{Korn}, 
than \li6 is destroyed by at least the same factor. This would yield an
upper limit close to \li6/\h1 $\simle 4\times 10^{-11}$, or 
\li6/\li7 $\simle 0.1$. However, since \li6 is more fragile than \li7 
it may, in principle, be destroyed by much larger factors than \li7,
as shown in a number
of stellar evolution studies~\cite{lideplete}.
A conservative limit of \li6/\li7 $\simle 0.66$ (corresponding approximately
to \li6 $\simle 2.7\times 10^{-10}$) was therefore applied in 
Ref.~\cite{jeda5}.

In Fig.~\ref{fig31} and Fig.~\ref{fig41}
constraints on decaying CHAMPs with intermediate life times
are presented. In both figures results are shown for a variety of hadronic
branching ratios $B_h$, with
Fig.~\ref{fig31} showing results for $M_x = 1\,$TeV, and Fig.~\ref{fig41}
for $M_x = 100\,$GeV.
In order to derive these constraints the rate for \h2(\he4$-X^- ,X^-){}^6$Li
as given in
Ref.~\cite{Hamaguchi} has been utilised.  
It is seen that for large hadronic branching
ratio ($B_h\simge 0.01$ for $M_x = 1\,$TeV and 
$B_h\simge 0.1$ for $M_x = 100\,$GeV)
constraints depend almost linearly on $B_h$
and are {\it not} different from those for neutral particles. This
seems somewhat surprising, due to the advocated 
power~\cite{Pospelov,Hamaguchi} of helium-CHAMP bound states to
produce \li6. Nevertheless, it is known that \li6 is also produced
abundantly by hadronic decays during that time~\cite{DEHS,jeda5}, and
this \li6 source is more important than that of 
\h2(\he4$-X^- ,X^-){}^6$Li at large $B_h$. 
The efficiency of \h2(\he4$-X^- ,X^-){}^6$Li has in any case been
overestimated in some papers~\cite{Hamaguchi,Kawasaki}. \li6 production
here is given by a convolution of two exponentials, (a) the
exponentially rising \he4$-X^-$ bound state fraction at $T\sim 10\,$keV,
and (b) the due to a Coulomb barrier exponentially decreasing
reaction rate. This leads to the bulk of the \li6 production in a very
narrow temperature interval $8\, {\rm keV}\simge T\simge 6\,{\rm keV}$.
Refs.~\cite{Hamaguchi} and ~\cite{Kawasaki} 
assumed the applicability
of the Saha equation for the \he4 bound state fraction (though earlier
studies Refs.~\cite{Pospelov,Kohri,Kaplinghat,Bird,Cyburt} did not). 
Numerical integration shows that the formation of appreciable
bound state fractions is slightly delayed when compared to the Saha
equation, since the recombination rate is of the same order as the
Hubble rate. This may be seen in Fig.~\ref{fig29}.
This slight difference results than in approximately one order of
magnitude less \li6 production, due to the convolution of exponentials.
Very recently, the same observation has also been made in Ref.~\cite{Steffen}
utilising the set of Boltzman equations relevant for \li6 production via
\he4$-X^-$ bound states as given in Ref.~\cite{Takayama}.
Conservative limits on CHAMPs in
the intermediate decay time interval are thus weaker
than initially thought.
Here the weakening is due to a lower
rate~\cite{Hamaguchi}, the failure of the Saha equation, and 
a too restrictive upper limit on the \li6 abundance.  
It is noted here that the photodisintegration of (\he4$-X^-$) bound
states as noted in Ref.~\cite{jeda6} is comparatively unimportant at small
$Y_x$, such that further weakening of the limit on CHAMPs does not result.
In any case, when $B_h\simle 10^{-2}$ bound state induced
production of \li6 becomes dominant over hadronic production. 
Since this is the case for, for example,
supersymmetric staus, bound state effects thus still remain very
constraining in particular scenarios~\cite{Pospelov}.

\section{Constraints on CHAMPs with long life times}

\bef
\epsfxsize=8.5cm
\epsffile[50 50 410 302]{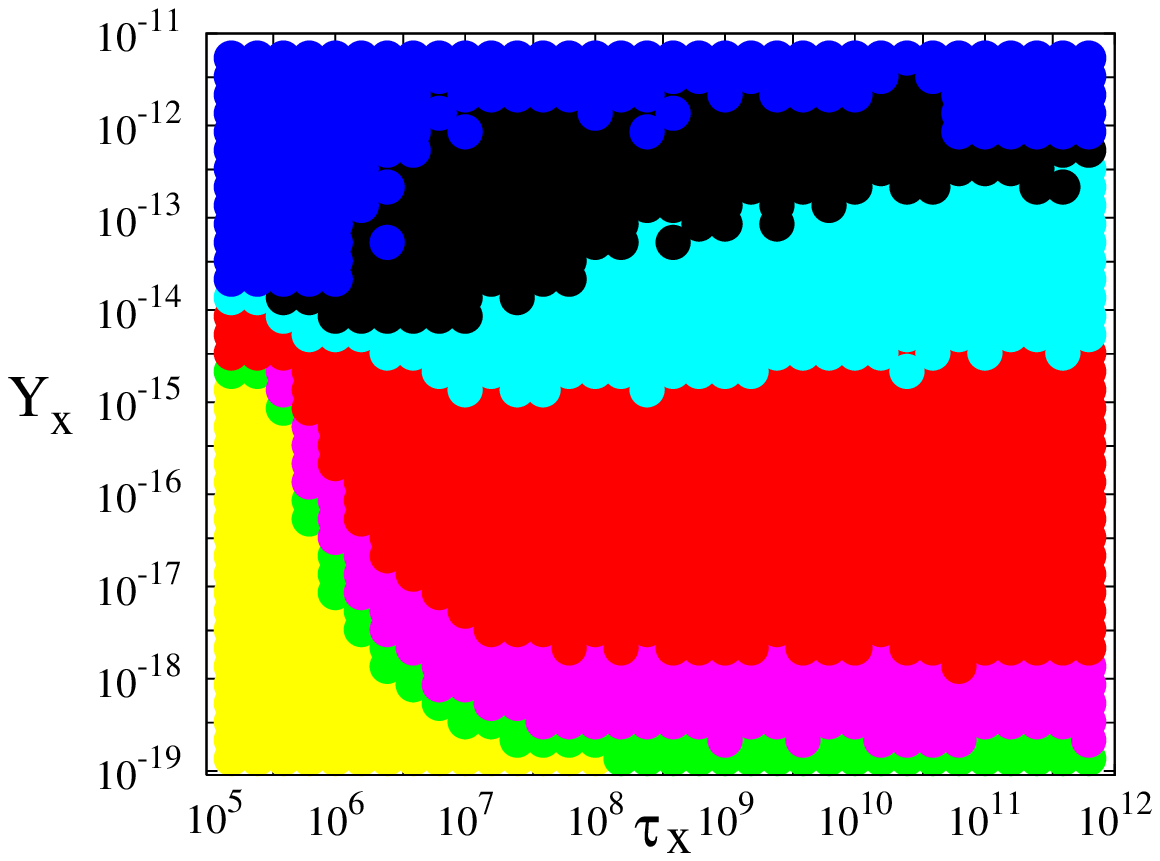}
\caption{Liklihood areas in the CHAMP-to-entropy ratio $Y_X$ -- CHAMP life time
$\tau_X$ (in seconds) plane
for bound-state BBN to obey the observational
constraints on light element abundances Eqs. 1. 
Results of a Monte-Carlo analysis varying
nineteen ill-determined reaction rates randomly
(see text for details) which significantly impact BBN yields
at late times $\tau_x\simge 10^6$sec are shown. From bottom to top, the
areas show liklihoods: $> 99\%$ lightest shade (yellow), 
$95\% - 99\%$ (green), $80\% - 95\%$ (purple), $20\% - 80\%$ (red),
$5\% - 20\%$ (light-blue), $1\% - 5\%$ (black), and $<1\%$   
(dark-blue), respectively, 
for CHAMP BBN to respect observational constraints. 
No effects of electromagnetic- and hadronic-
cascades due to the CHAMP decay have been taken into account
(cf. Fig.~\ref{fig33}). The Monte-Carlo analysis presented in this
figure employs $\sim 1.5\times 10^6$ independent BBN calculations.} 
\label{fig32}
\eef

\bef
\epsfxsize=8.5cm
\epsffile[50 50 410 302]{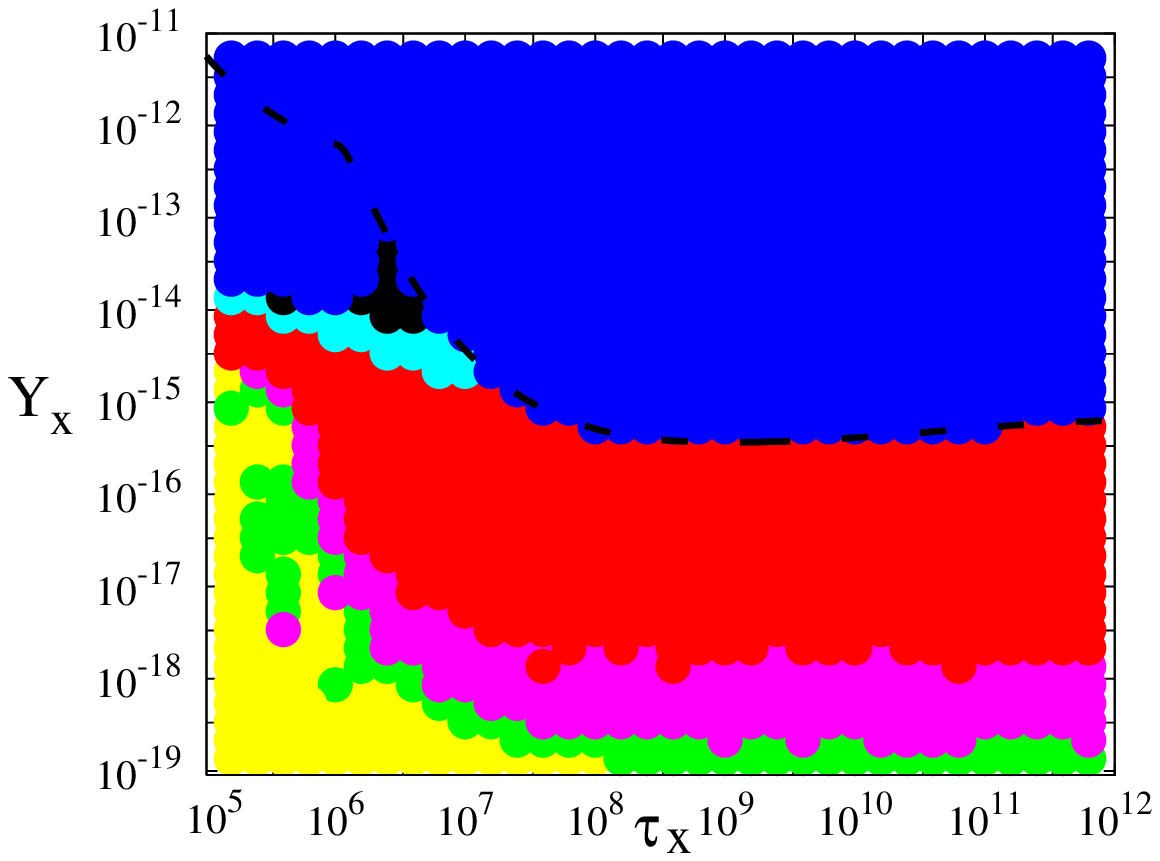}
\caption{As Fig.~\ref{fig32} but including the effects of electromagnetic
cascades during $X$-decay,
assuming that a fraction $f_{EM} = 1$
of the particles rest mass (taken $m_X = 100\,$GeV) is
converted into electromagnetically interacting particles. The hadronic
branching ratio was set to $B_h = 0$. The dashed line shows the analogous
limit for neutral relics.}
\label{fig33}
\eef

\bef
\epsfxsize=8.5cm
\epsffile[50 50 410 302]{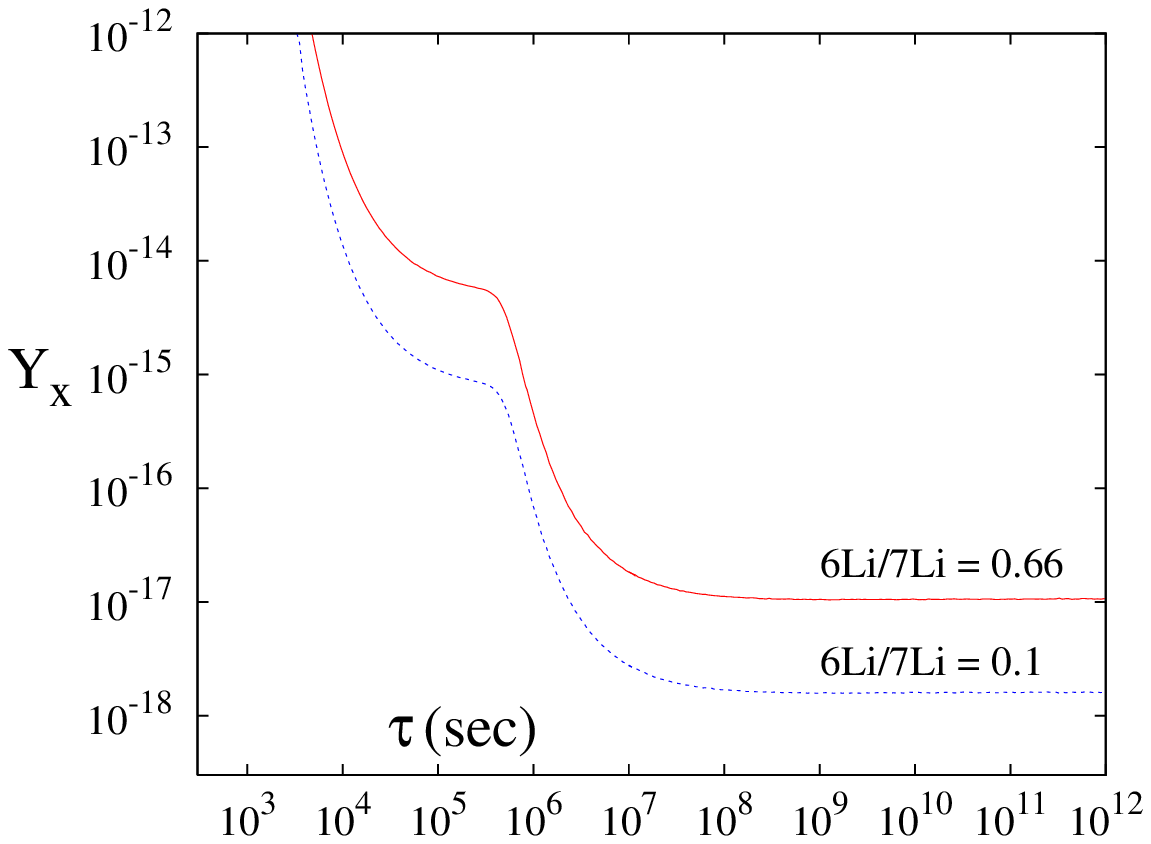}
\caption{Isocontours of \li6/\li7 $=0.66$ solid (red), and
\li6/\li7 $=0.1$ dashed (blue), in the CHAMP-to-entropy $Y_x$ - CHAMP
life time $\tau_x$ plane for rates determined in the unreliable
Born approximation~\cite{jeda6}. 
The present figure is {\it not} supposed to be utilised
for limiting CHAMP abundances. It rather is for illustrative purpose,
showing the possible importance of the \he4 + (\h2$-X^-) \to X^- + {}^6$Li
reaction to further tighten limits on long-lived CHAMPS by orders of
magnitude due to late-time \li6 production. A hadronic branching ratio of
$B_h = 0$ has been assumed.}
\label{fig34}
\eef

When the life times exceeds 
$\tau_x\simge 5\times 10^5$sec it becomes substantially
more difficult to place reliable limits. This is due to a large number
of CHAMP-induced 
reactions becoming important at $T\simle 1\,$keV, in particular all
Coulomb-unsupressed
nuclear reactions shown in Table II and 
Table III of Ref.~\cite{jeda6}, as well
as the CHAMP exchange reactions shown in Table IV of that paper. 
This comprises a total number
of nineteen reactions.
Though all rates
have been determined numerically in the Born approximation in Ref.~\cite{jeda6},
as the Born approximation is likely to fail badly,
results become uncertain. In order to still arrive at a reliable
result one is thus
forced to perform a Monte-Carlo analysis, varying all ill-determined 
reaction rates within conservative ranges. This has been done in the
present paper. In particular, the Born approximation values of the
rates given in Ref.~\cite{jeda6} (shown in Figs. 3 and 4, 
as well as in Table~III and~IV of that paper),
have been taken as benchmarks. For each reaction a random generator
determined a factor $f_i$ with which the benchmark rate was multiplied.
These factors where generated with a probability distribution flat
in logarithmic space, and between values 
$1/f_i^{cut}\leq f_i \leq f_i^{cut}$. For the reaction-rate dependent
conservatively chosen $f_i^{cut}$ the reader is referred to 
Table~VII of Ref.~\cite{jeda6}. For each point in parameter space, i.e. for 
$Y_x$ and $\tau_x$, this procedure was repeated a $1000$ times
in order to arrive with one thousand different randomly chosen sets
for the 19 ill-determined reaction rates. For each realization of
reaction rates an indpendent BBN calculation was then performed
and compared to the observational constraints.

One may wonder if 1000 realizations for the reaction rates are
actually sufficient for sampling the, a priori, complicated
probability space. After all even only adopting two values for each
reaction rate, a large rate and a small rate, already yields 
$2^{19}\approx 5\times 10^5$ different possibilities for sets of
reaction rates. For a few
individual points in $Y_x$ and $\tau_x$, 
I have therefore generated $10^5$ different realizations of rate
combinations. Comparison with the results by only a 1000 realizations
shows that the simulation with a 1000 realizations may be trusted
approximately to the 1\% level (i.e. $< 10$ BBN runs passing or failing
observational constraints). This is true since at an individual point 
in the
$Y_x$-$\tau_x$ plane, results mostly only depend on a number
$\sim 4-6$ of rates, with all other rates being less important.
Which rates are most important for the BBN yields, however,
depends on the location in the $Y_x$-$\tau_x$ plane.  
For example, at large $Y_x$ (and $\tau_x$) results depend sensitively
on the reactions: \h3(\he4$-X^- ,X^-){}^7$Li,
\li6(\h1$-X^- ,X^-)$\he4 + \he3, and 
(\h1-$X^-$)+ \h2$\to$\h1 + (\h2-$X^-$) (and to a lesser 
degree \be7(\h1$-X^- ,X^-){}^8$B, \he4(\h3$-X^- ,X^-){}^7$Li, and 
(\h1-$X^-$)+ \he4$\to$\h1 + (\he4-$X^-$)),  
whereas for small $Y_x$ (still large $\tau_x$)
they depend mostly on reactions: \he4(\h2$-X^- ,X^-){}^6$Li, 
(\h1-$X^-$)+ \h2$\to$\h1 + (\h2-$X^-$) and 
(\h1-$X^-$)+ \he4$\to$\h1 + (\he4-$X^-$) (and to a lesser degree
(\h2-$X^-$)+ \he4$\to$\h2 + (\he4-$X^-$)).

The results of the present Monte-Carlo analysis are shown in 
Figs.~\ref{fig32} and~\ref{fig33}. Shown are the liklihood areas
in the $Y_X$-$\tau_X$ plane by different shading (coloring)
that $> 99\%$, $95\% - 99\%$,$80\% - 95\%$,$20\% - 80\%$,
$5\% - 20\%$,$1\% - 5\%$, and $< 1\%$ of all randomly
generated models at the same $Y_X$ and $\tau_X$ obey the observational
constraints. Note that the CHAMP-to-entropy ratio $Y_X$ is easily
converted to $\Omega_X$, the fractional contribution of the
$X$-particle to the present critical density, if it had not decayed,
via $\Omega_X h^2= 2.73\times 10^{11}\,Y_x\,(M_x/1\, {\rm TeV})$,
where $h$ is the present Hubble constant in units 
100 ${\rm km s^{-1} Mpc^{-1}}$. Note also, that in all constraint
figures $Y_x$ denotes the
total CHAMP-to-entropy ratio, assuming that only half of all CHAMPs
are negatively charged.
Whereas Fig.~\ref{fig32} shows
results when the CHAMP decay is not associated with electromagnetic-
or hadronic- energy release, Fig.~\ref{fig33} assumes a $100\%$ 
electromagnetic CHAMP decay and the associated electromagnetic
cascade nucleosynthesis. 
It is evident from Fig.~\ref{fig32} that, when only effects of bound
states are taken into account but not injection of energy during the
decay, the probability distribution is extremely flat. In particular,
whereas one still finds a small fraction of
models which fail for $Y_x$ as low as
$\sim 10^{-19}$ only at $Y_x \simge 10^{-12}$ one may exclude CHAMP BBN
at the $> 99\%$ confidence level. Conservatively, and in the absence
of more reliable rates, even $Y_x$ as large as $10^{-12}$ may thus not
reliably be ruled out. This is five orders of magnitude less stringent
than the initial claim~\cite{Pospelov}.
Note, however, that such large $Y_x$ may only
be (conservatively) acceptable for invisible, mass-degenerate, or
stable CHAMPs, associated with very little (or no) injection of
electromagnetically interacting energy (i.e. high-energy $\gamma$'s
and $e^{\pm}$'s). This may be seen from Fig.~\ref{fig33}, where for
larger life times $Y_x \simge 10^{-15}$ is ruled out at the $> 99\%$
level. Here models are ruled out principally due to violating the \he3/\h2
upper limit due to \he4 photodisintegration. 
This fact is essentially not changed by the $X$-particle being
charged, as may be seen by inspection of the results in 
Ref.~\cite{jeda5} for neutral decaying
particles or by the dashed lines in Fig.~\ref{fig33}, which shows the
analogous limit for neutral particles. 
Immediately below the constraint line already
between $20 - 80\%$ of all models yield acceptable abundance yields.
This parameter space, is thus, currently not ruled out. However, it
is conceivable that constraints on CHAMPs may be significantly tighted
in future, to values possibly as low as $Y_X\simle 10^{-17}-10^{-18}$,
in case rates for the important reactions 
\he4(\h2$-X^- ,X^-){}^6$Li, (\h1$-X^-$) + \h2$\to$ (\h2$-X^-$) + \h1,
and (\h1$-X^-$) + \he4$\to$ (\he4$-X^-$) are determined, and contrive
to yield unacceptably large \li6 at low $Y_X$. 
Here it is noted that the by far most
important \li6 producing reaction, potentially leading
to such stringent constraints, is \he4(\h2$-X^- ,X^-){}^6$Li rather
than \h2(\he4$-X^- ,X^-){}^6$Li, yielding the bulk of the \li6 at temperatures
$T\simle 1\,$keV and not $T\approx 6-8\,$keV. Important here is the
CHAMP exchange reaction (\h1$-X^-$) + \h2$\to$ (\h2$-X^-$) + \h1 which
may continously produce (\h2$-X^-$) bound states.
(cf. Ref.~\cite{jeda6} for more detail). 
This trend may be seen in Fig.~\ref{fig34}, where \li6/\li7 isocontours 
are shown for a CHAMP with $B_h = 0$, and when the Born approximation for
all rates is utilised. It is seen that for $\tau_x\simge 10^6$sec factor
$\sim 1000$ more \li6 is synthezised than for $\tau_x\simle 10^6$sec. 

It is evident from Figs.~\ref{fig32} and~\ref{fig33} that the probability
distributions only flattens significantly for CHAMP life times 
$\tau_x\simge 5\times 10^5$sec. 
For $10^5{\rm sec}\,\simle\tau_x\simle\, 5\times 10^5$sec
the probability passes from $> 99\%$ observationally acceptable models
to $< 1\%$ observationally acceptable models within one decade of $Y_x$.
This is because for shorter $\tau_x$ possible
late-time destruction of \li6 and \li7
may not yet be
efficient due to large (\h1$-X^-$) bound state fractions only
forming at $T\simle 1\,$ keV. Up to $\tau_x\simle 5\times 10^5$sec 
one may therefore
approximately use the bounds on CHAMPS as given in Section 2. When
$\tau_x\simge 5\times 10^5$sec late-time processing (destruction)
of \li6, \li7, and \h2 may
occur efficiently. For long life times it is thus
proposed to use {\it only}
the constraint imposed by possible \he3/\h2 overproduction. This constraint
is not affected by bound states and the uncertainties in reactions including
bound states. The \he3 isotope is special in that it may not be destroyed
without destroying \h2 as well. This is because the reaction
\he3(\h1$-X^- ,X^-){}^4$Li is endothermic and other reactions not involving
\h2 nuclei are Coulomb supressed. It is thus conservative to use the
\he3/\h2 constraint, in particular, since there are other \h2 destroying
(but not producing) reactions within bound state nucleosynthesis. This
translates into using the constraints given in Ref.~\cite{jeda5} for $B_h =0$
(and shown by the dashed curve for $\tau_x\simge 10^6$sec in 
Fig.~\ref{fig33}).
Since the fraction of rest mass $f_{EM}$ which is converted to
electromagnetic interacting energy has been assumed $f_{EM}=1$ 
for $B_h = 0$ in the figures of Ref.~\cite{jeda5}, the constraint has to be
rescaled accordingly, when neutrino losses are significant, or 
close-to mass degeneracy between mother and daughter particle exists.
Finally, in the window
$5\times 10^5{\rm sec}\simle \tau_x\simle 
3\times 10^6{\rm sec}$ one may apply an additional constraint, stronger than
that due to \he3/\h2 overproduction. From Figs.~\ref{fig32} and ~\ref{fig33} 
it is found that for
$Y_x\simge 3\times 10^{-14} - 10^{-13}$
only $\simle 1\%$ of all models are observationally acceptable.

A procedure very similar to this has been very recently applied by 
Ref.~\cite{Steffen} to derive a lower limit on the gaugino mass parameter
$m_{1/2}$ in the constrained minimal supersymmetric standard model (CMSSM)
when the gravitino is the lightest supersymmetric particle (LSP).
This study considered only the 
catalyzed reaction \h2 + (\he4-$X^-$) $\to$ \li6 $+X^-$ for
intermediate life times of the next-to-LSP (NLSP), arguing that even
without the \li6 bound long NLSP life times were already ruled out
priorly by the \he3/\h2 upper limit. Since in the parameter space 
under investigation the NLSP is the stau, which has small hadronic
branching ratio, the conclusion for
$\tau_x\simle 5\times 10^5{\rm sec}$
is thus not expected to change much when the present results are used. 
Using the results of the Monte-Carlo analysis shown in Figs.~\ref{fig32}
and \ref{fig33} 
typical stau-to-entropy ratios 
$Y_{\tilde{\tau}}\simge 10^{-13}$ should be also
disallowed for 
$5\times 10^5{\rm sec}\simle \tau_x\simle 3\times 10^6{\rm sec}$ due
to \li6 overproduction and for $\tau_x\simge 3\times 10^6{\rm sec}$
due to \he3/\h2 overproduction, rendering the conclusions of 
Ref.~\cite{Steffen} likely unchanged.

\vskip 0.15in

\section{Conclusions}

In this letter the results of
a detailed study of constraints on charged massive particles
$X^-$ 
from Big Bang nucleosynthesis was presented. It was pointed out earlier
that bound states between \he4 and negatively charged $X^-$ may lead to 
the efficient catalytic production of \li6 at $T\approx
6-8\,$keV~\cite{Pospelov}. Recently I have shown~\cite{jeda6}, that
BBN with CHAMPs enters a second phase of nucleosynthesis at $T\simle 1\,$keV,
capable of destroying all priorly synthesized \li6. Altogether
nineteen reactions important for late-time BBN were identified. When these
processes are included in the analysis,
drastic changes concerning limits on the existence of CHAMPs
when compared to those priorly 
derived~\cite{Pospelov,Hamaguchi,Cyburt,Kawasaki,Pradler}, are obtained.

Limits on the existence of CHAMPs during and after BBN are derived
in two different decay time ranges. For very short life times
$\tau_x\simle 3\times 10^2$sec limits are independent of the decaying
particle being charged, or not.
In the range 
$3\times 10^2{\rm sec}\simle\tau_x\simle 5\times 10^5{\rm sec}$ 
the important rates
(i.e. \h2(\he4$-X^- ,X^-){}^6$Li) are relatively well 
determined~\cite{Hamaguchi}, such that a Monte-Carlo analysis
may be avoided. It is stressed that the approximation of the \he4-bound
state fraction by the Saha equation leads to a factor $\sim 10$
overestimate in the synthesized \li6 abundance. 
Due to
substantial reaction rate uncertainties a full Monte-Carlo 
analysis had to be performed to obtain reliable and conservative
bounds in the decay time range 
$5\times 10^5{\rm sec}\simle\tau_x\simle 10^{12}{\rm sec}$.
It was found that when a number of reaction rates are large, and when 
electromagnetic energy injection is absent, baryon-to-entropy ratios
$Y_x$ as large as $\simge 10^{-12}$ may be observationally acceptable.
On the other hand, in case a number of reaction rates are
determined more precisely, in particular the rates for
\he4(\h2$-X^- ,X^-){}^6$Li, (\h1-$X^-$) + \h2$\to$ (\h2-$X^-$) + \h1,
and (\h1-$X^-$) + \he4$\to$ (\h1-$X^-$) + \he4, it may be conceivable that
limits on long-lived CHAMPs are improved by orders of magnitude, with
the CHAMP-to-entropy ratio possibly constrained to be below
$10^{-17}-10^{-18}$. A prescription is given for how to place conservative
limits on CHAMPs, given current reaction rate uncertainties.

The final result of this study is somewhat surprising. When electromagnetic-
and hadronic- energy release are included, and within the reaction
rate uncertainties, conservative limits on charged decaying
particles are no stronger than those on neutral particles. The exception
here is the decay time range $10^3-3\times 10^6$sec but only
when the hadronic branching ratio is small $B_h\simle 10^{-2}-10^{-1}$, as
is the case for supersymmetric staus.
\vskip 0.1in

\vskip 0.1in
I acknowledge helpful discussions with 
and M.~Asplund, S.~Bailly, O.~Kartavtsev, K.~Kohri, 
A.~Korn, G.~Moultaka, M.~Pospelov, J.~Rafelski,
G.~Starkman, F.~Steffen, V.~Tatischeff, and T.~Yanagida. 
\vskip 0.1in

\section{Erratum:
Bounds on long-lived
charged massive particles from Big Bang nucleosynthesis}

\bef
\epsfxsize=8.5cm
\epsffile[50 50 410 302]{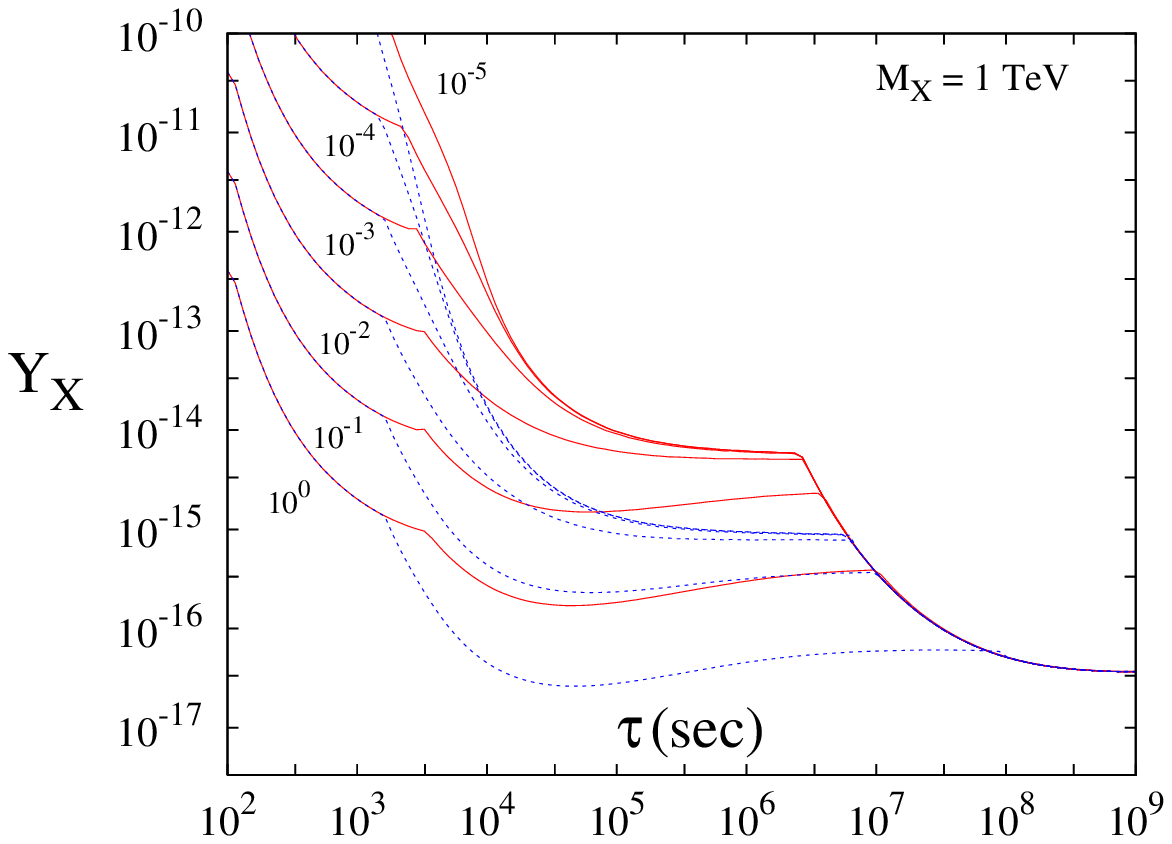}
\caption{Limits on the primordial CHAMP-to-entropy ratio 
$Y_x = n_{X}/s$ (with $n_{X^-}/s = Y_x/2$) as a function of
their life time $\tau_x$.
Shown are constraint
lines for CHAMPs of mass $M_x =1\,$TeV and a variety
of hadronic branching ratios $B_h = 10^{-5}-1$,
as labeled in the figure. Solid (red)
lines correspond to the conservative limit \li6/\li7 $< 0.66$, 
whereas dashed (blue) lines correspond to
\li6/\li7 $< 0.1$. It is seen that only for CHAMPs with $B_h\simle 10^{-2}$
the effects of bound states become important. For small decay times
$\tau_x\simle 10^3$sec and large decay times $\tau_x\simge 10^7$sec 
the limits on CHAMP abundances are virtually identical to those
on the abundance of neutral relic decaying particles~\cite{jeda5}.}
\label{fig90}
\eef

\bef
\epsfxsize=8.5cm
\epsffile[50 50 410 302]{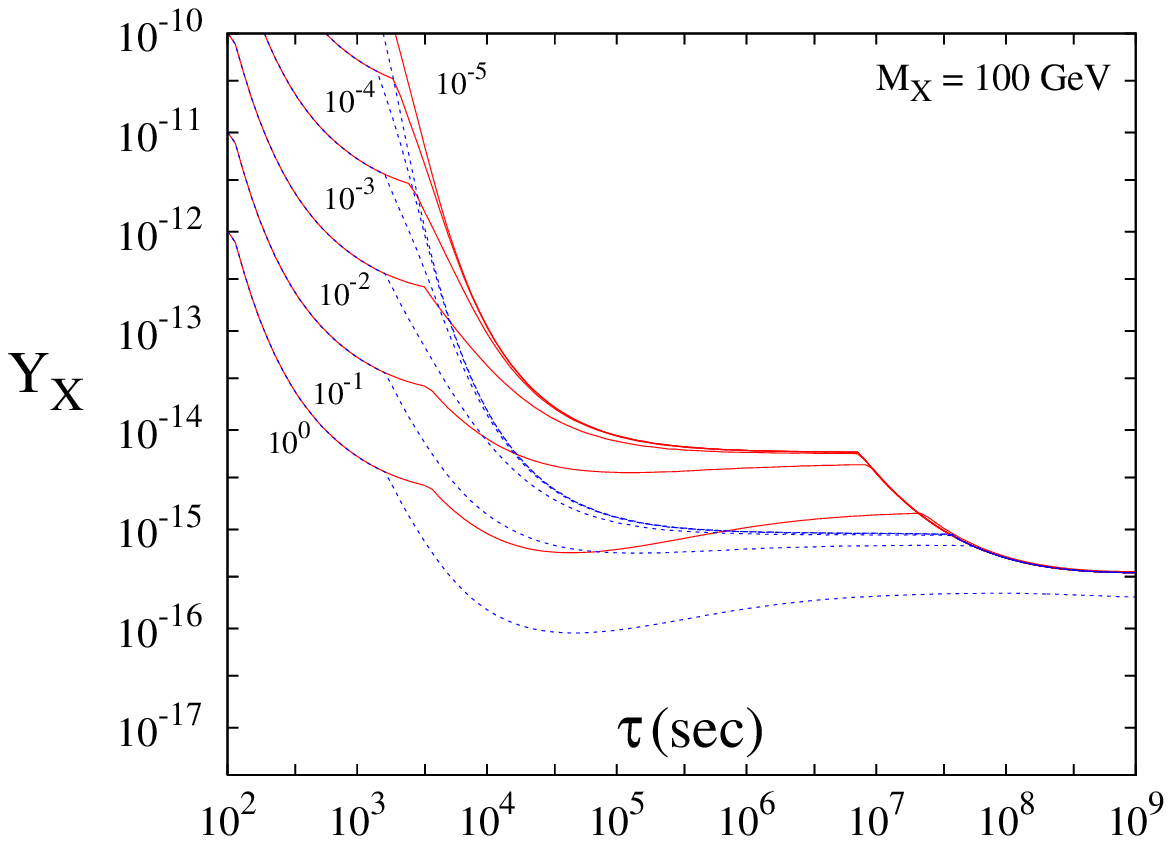}
\caption{As Fig.~\ref{fig90} but for $M_x = 100\,$GeV.}
\label{fig91}
\eef

A recent detailed computation by 
Kamimura {\it et al.}\cite{Kamimura:2008fx}
of the most important reaction rates
entering bound state nucleosynthesis 
(see also the discussion in Ref.~\cite{Pospelov:2008ta}) has
established that late time $\tau\simge 10^6$sec processing of
light element abundances as envisioned possible by the present
author~\cite{jeda6} does not usually take place.
This is mostly because of the abundance of proton-$X^-$ bound states 
staying small due to
efficient exchange reactions transferring $X^-$ from $p$ to \he4, but also
due to Coulomb barriers between $p$ and nuclei being only partially
shielded when protons are in bound states. The Monte Carlo analysis as
performed in Section II and III is therefore superfluous. Figs. 4-6 of the
paper are replaced by Figs. 7 and 8 in this erratum. It is cautioned, however,
that for relatively large $Y_X > 2 Y_{\rm {}^4He}$ late time processing
may still occur to some degree.


\begin{thebibliography}{99}
\bibitem{remark301}
The effects of annihilation of $X^-$ with $X^+$ in $X^--X^+$ bound states 
forming to some degree~\cite{Bird} 
are subdominant compared to those of the decay
itself.
\end{thebibliography}
\end{document}